# Keyboard Surface Interaction: Making the keyboard into a pointing device

Julian Ramos*, Zhen Li†, Johana Rosas‡, Nikola Banovic*, Jennifer Mankoff *, Anind Dey*

* Human Computer Interaction Institute, Carnegie Mellon University, Pittsburgh, USA
† Tsinghua University, Beijing, China
‡ Psychology Department, University of Pittsburgh, Pittsburgh, USA

**ABSTRACT**
Pointing devices that reside on the keyboard can reduce the overall time needed to perform mixed pointing and typing tasks, since the hand of the user does not have to reach for the pointing device. However, previous implementations of this kind of device have a higher movement time compared to the mouse and trackpad due to large error rate, low speed and spatial resolution. In this paper we introduce Keyboard Surface Interaction (KSI), an interaction approach that turns the surface of a keyboard into an interaction surface and allows users to rest their hands on the keyboard at all times to minimize fatigue. We developed a proof-of-concept implementation, Fingers, which we optimized over a series of studies. Finally, we evaluated Fingers against the mouse and trackpad in a user study with 25 participants on a Fitt's law test style, mixed typing and pointing task. Results showed that for users with more exposure to KSI, our KSI device had better performance (reduced movement and homing time) and reduced discomfort compared to the trackpad. When compared to the mouse, KSI had reduced homing time and reduced discomfort, but increased movement time. This interaction approach is not only a new way to capitalize on the space on top of the keyboard, but also a call to innovate and think beyond the touchscreen, touchpad, and mouse as our main pointing devices. The results of our studies serve as a specification for future KSI devices, (independent of sensing technology).

**ACM Classification Keywords**
H.5.2. Information interfaces and presentation: User Interfaces: Input devices and strategies.

**INTRODUCTION**
The computer mouse, introduced in 1964 by Douglas Engelbart, is the preferred pointing device for computer users. However, it is by no means perfect. In our own study, we found that the main complaint from using a mouse for 22 out of 25 participants was discomfort and fatigue in different parts of the arm caused by having to move the hand between the mouse and the keyboard while performing mixed pointing and typing tasks. Also, manipulation of the mouse has shown to create additional discomfort in the hand due to wrist extension, and frequent mouse usage has been found to be a risk factor for Carpal Tunnel Syndrome (CTS) and repetitive hand strain injuries [1,9]. In contrast, people who spend more time using the keyboard than the mouse at work, have significantly lower risk for CTS [1]. Several technologies have been developed that could potentially reduce this discomfort by eliminating the use of an interaction device like the mouse or trackball. Examples of technologies that do not use an interaction device include touch sensing technologies like the touchpad on laptops, touchscreens on smartphones and tablets, and mid-air interaction (*e.g.*, Leap Motion, Kinect, Senz3D).

The main limitation of these alternative technologies in becoming everyday pointing devices is poor performance. In this paper, we will refer to performance as the combination of movement time, homing time, and error rate. Movement time [27] refers to the time elapsed from the moment the user starts moving the pointer on the screen to when she clicks on a target. High movement time is undesirable because it implies that more time is required to click on a target making the whole action more inefficient. Homing time [4] is defined as the time elapsed between when a pointing target is shown on a screen and when movement of the pointer is detected. Homing time is composed of the reaction time of the user and the time it takes the hand to physically travel to reach the pointing device. Thus, a high homing time implies more time is needed to reach the pointing device and results in longer time being required to complete a pointing task making the whole task inefficient. Error rate refers to the quantity of errors made by the user, where an error is defined as clicking outside (missing) the target at least once.

For mid-air interaction devices, low performance is explained by low resolution or low sampling rates. For example, the Kinect, has a sampling rate of 30 samples per second, leading to a perceivable lag in responsiveness, and higher movement time. In contrast, modern mice and touchpads have sampling rates between 100 to 1000 samples per second, with no perceivable lag. Additionally, mid-air and touchscreen devices cause a feeling of heaviness in the arm [14] even for interactions with durations shorter than 5 minutes [3]. Touchscreen devices (and mid air devices depending on the position of the screen), suffer from occlusion (users cannot see what they are pointing at) and the fat finger problem (users click on more than one target at a time, when targets are smaller than their fingers). More generally, mid-air interaction and touchscreens have failed to become mainstream pointing devices for laptop and desktop computers due to poor performance and increased discomfort.

While the mainstream pointing devices, touchpad and mouse, tend to have better performance, they also suffer from increased discomfort. This is due partly to wrist extension issues (bad for both devices but significantly worse for the touchpad [26]), and partly to having to move one's hand between the keyboard and the touchpad or mouse during mixed typing/pointing interactions. The distance between the keyboard and touchpad or mouse also leads to reduced performance in the form of increased homing time.

To address the performance and discomfort concerns of these existing pointing technologies, we introduce keyboard-surface interaction (KSI). KSI is a technique that allows the user to perform pointing tasks on the surface of the keyboard itself, using the user's hand as the pointing device. Our system leverages the property of physical keyboards, which allows the user to rest her fingers on the keys without pressing them. Allowing the user to rest her hands on the keyboard reduces homing time and decreases perceived discomfort. Unlike previous similar approaches [10], we successfully demonstrate through a rigorous evaluation that a KSI device can be on par with commercially mature technologies, this is our main contribution.

To demonstrate the feasibility and value of KSI we built a proof of concept sensing system that we named *Fingers*. We optimized Fingers along several dimensions (sensing technology; relative *vs.* absolute motion; and invocation method), using results from a series of Fitts' law style studies to refine the prototype along these dimensions. The lessons learned from these iterations represent a second contribution of our work.

The final contribution of our paper is a study demonstrating the value of KSI. We compare Fingers against mainstream pointing devices (trackpad and mouse) on a mixed typing and pointing task using a Fitts' Law test. We chose a mixed typing/pointing task because it is more representative of daily computing tasks than a pointing-based Fitts' Law test alone. We hypothesize that KSI is more comfortable because it allows the hand to rest in a natural position on the keyboard and reduces to a minimum the effort to switch between pointing and typing tasks. Our study provides evidence for this hypothesis: 1) for more experienced Fingers users, we found that Fingers decreases discomfort in comparison to both the trackpad and mouse. 2) Fingers has a comparable error rate and shorter homing time than the mouse and the trackpad. The results of our studies serve as a specification for future KSI devices, (independent of sensing technology) with respect to error rate, speed, and spatial resolution.

In the following section we provide an overview of different pointing devices and their relation to discomfort, fatigue and injuries. We then introduce the motivation for Keyboard Surface Interaction, and present the different prototypes we implemented together with a set of three user studies to evaluate and optimize the prototypes' performance. Then, we describe our final user study that compared Fingers, to the mouse and trackpad, and show that for more experienced users of Fingers, Fingers outperforms the trackpad in performance and comfort. While it has mixed performance ratings compared to the mouse, it also results in less discomfort. We end with a discussion of future work and our main conclusions.

**RELATED WORK**

Since the introduction of the mouse, most of its variations have retained the same idea: the user's hand manipulates an object, and this object's movement is measured and used to control the position of the pointer on a computer's screen. The touchpad is the most common mouse alternative for laptop computers, and touchscreens have become the norm for smart phones and tablet devices. In this section, we examine other options for controlling a pointer on a screen.

Different parts of the body such as the head [25,32], eyes [32], mouth [32] and feet [32] have also been explored as pointing devices. Nonetheless, they have been popular only among people with disabilities or that suffer from injuries like carpal tunnel syndrome. There is no particular advantage for typical users to use any of these devices. Moreover, a user's performance is decreased when using these alternatives due to the increased time and effort to complete a pointing and typing task, which leads to a less satisfactory user experience.

Another approach to replace the mouse and touchpad is to make the user's hand the pointing device, an example of this kind of device is Touch&Type [10], a touch sensitive keyboard. The keys themselves are touch sensitive and a single finger or the entire hand can be used to steer the cursor. Although interesting, the evaluation in [10] is unconventional for understanding performance. There is no Fitts' Law test, and the average of best times over sessions used to evaluate final performance, may have skewed results towards outliers with the best performance. While the device, the authors claim, is close in performance to the touchpad, from the results section, it's not clear that this is a significant result. Basic details are missing for replication including screen resolution, target size, number of participants and error rate. A similar device is presented in [22]. In this work, users were required to move their hand adjacent to the keyboard and to imitate the operation of a regular mouse. More explicitly, the method required a user to "click" by moving his index finger up and down despite the absence of any mouse. Results on the error rate, discomfort or acceptance of this approach were not reported.

Another similar device/technique is FlowMouse [31]. FlowMouse uses a 30Hz gray scale camera with a resolution of 640x480. The camera is used to detect the general flow of motion in its field of view to determine how to move the pointer. FlowMouse uses a button on top of the left mouse button for clutching (turning on and off the device). An advantage of FlowMouse is that it does not rely on detecting the hand and instead moves the pointer based on general movement in the camera's field of view; however an evaluation shows that FlowMouse was significantly slower in general than the trackpad. Another significant disadvantage of this technique is that it will be sensitive to the movement of the non-pointing hand or arm movements, as both introduce noise into the system.

Mid-air and touch screen devices also can make the user's hand a pointing device. Mid-air interaction, which occurs above the keyboard, leads to a feeling of heaviness and fatigue [3,14] in the arms, making this method useful only for short-term use. Touch-screen devices are also used as alternatives to the mouse however occlusion and the fat finger problem [24] make them a good option mostly for non-productivity related tasks, kiosk terminals [3] and single point interactions [11].

Other efforts to replace the mouse are concepts that fuse the mouse and keyboard. However, all of them are only reported in patent filings, and none of them have been commercialized or evaluated. One such example is the "computer keyboard pointing device" [5] that uses a small cross-shaped touch sensing device that is inserted in between the keys Y,U,H and J on a QWERTY keyboard. This concept requires the user to perform pointing tasks on a very narrow and specific area of the keyboard by continually swiping her finger. Only horizontal or vertical movements can be performed, and not diagonal ones. This implies that the movement of the pointer is not smooth and instead has a taxicab geometry [19]; *i.e.*, movements are only vertical or horizontal. Similarly, in the "keyboard pointing device" patent [21], the keyboard keys control the pointer. This accessibility function is already mainstream on all major computer operating systems, however it is generally aimed at people with motor control impairments who cannot use the mouse or touchpad [33]. More recently, Apple filed a patent application [8] in which the idea of a keyboard with touch sensitive capabilities was described. While in theory the device could be used for recognizing gestures and perform pointing tasks there was no discussion of the performance of the device. Many other patents were found [7,13,16,20] with similar approaches, however, to our knowledge, the corresponding devices have not been made available to the public nor is there any reference to their performance or implementation.

More recently, some researchers have tried to enhance the functionality of the keyboard by including some touch [2] and proximity sensing capabilities [29]. However, their goal was not to create an alternative to existing pointing devices, but to make the keyboard into a gesture sensing device. Nonetheless, the low spatial sensing resolution of those devices may not be accurate enough for pointing tasks. Similarly, in [30], although the authors hint towards the introduction of a pointing device there is no evaluation of pointing performance. Their device uses spatial resolution of 58x20, with sampling rate of 20Hz, but our own tests show that at least 90Hz is needed to avoid lag during pointing tasks, making the device inadequate for pointing. The paper proposes a mode switching prediction layer, but missing details of their evaluation make it unconvincing. The implementation uses motion history images that require several frames (number of frames isn't specified). More than 3 will make their mode switching slower than our approach.

### Discomfort, fatigue and injuries

Different research studies have found positive associations between wrist and hand pain and discomfort, and duration of mouse use [15]. Similarly, [1] presented the relation between the frequency of keyboard use at work and prevalence of CTS. Results showed that participants, who spend more time using the keyboard than the mouse at work, had significantly lower risk for CTS.

Thus, it is our hypothesis that pointing tasks performed directly on the keyboard may result in less discomfort, but possibly at the cost of worse performance for pointing tasks. This idea was partially tested by Douglas [6] (discomfort or fatigue was not a metric recorded or used in her study), by comparing the performance of an isometric joystick (positioned under the 'J' key of a keyboard), to a mouse and trackpad, on a mixed pointing and typing task. The mouse and trackpad performed better despite the joystick's significantly lower homing time. Although the isometric joystick decreased homing time, it increased movement time and made the entire pointing and typing task slower compared to the mouse. No measures of discomfort, fatigue or error rate were reported [6]. These results suggest that homing time can be reduced by having the pointing device located directly on the keyboard. If such a device also had low movement time and a low error rate, it could be competitive with the touchpad and mouse.

This previous research is the motivation for our own research: to address issues of both discomfort and performance for everyday tasks, by creating a pointing device on the surface of the keyboard.

### KEYBOARD-SURFACE INTERACTION

To decrease the discomfort caused by frequent switching between pointing and typing devices common in everyday computing tasks, we introduce keyboard-surface interaction (KSI). In our novel input technique, the user performs pointing tasks while resting her hands on the keyboard's surface. The user's hand is the pointing device, and the surface of the keyboard is the sensing area in which the system detects the user's pointing.

The main tenets of KSI are the next:

1. *The user hands should always rest on the keyboard's surface*: The user can and should keep her hands resting on the keyboard surface in order to perform a pointing task. This helps to minimize hand movement (which we will show was an issue with mouse users in our comparative study of KSI) and may decrease risk for CTS by eliminating the need to use a mouse.
2. *Supports whole keyboard interaction:* While in our studies we only tried out a single area of the keyboard a KSI device should be able to sense any area of the keyboard. This property will allow a KSI device to move its sensing area to adjust for the users needs. Moreover a KSI device may allow the user to perform a pointing task with one hand while allowing for other kinds of interactions with the other hand.

KSI has two modes: 1) typing mode, and 2) pointing mode. In typing mode, KSI does not recognize any hand movement as pointing. This mode was created to avoid unintentional pointing while typing. In pointing mode, KSI gives the user full control over the movement of the pointer on the screen. A system can achieve this by tracking one of the fingers or a specific part of the hand and mapping this movement to the pointer movement.

While the tenets for KSI are straightforward, there are multiple ways in which it could be implemented. In this section, we present a proof of concept sensing system, that we call Fingers. Fingers was developed in an iterative fashion, using a series of Fitts' law studies and prototypes to optimize its performance. Our iterative design process focused on improving error rate, homing time, movement time, and comfort.

Our results are tightly related to the specifications of our sensing technology used in our proof of concept implementation. These specifications should be taken as the minimal requirements to obtain similar results even with a different sensing technology.

Here we describe the general concept of Fingers, present different variations of Fingers that we built and tested, and discuss lessons learned to improve its performance.

**Overview of Fingers**
The basic concept behind our Fingers prototype is hand tracking. However, hand tracking can be difficult due to the multiple shapes that the hand can take and how quickly it can move. Moreover, the fastest marker-less hand tracking systems (Kinect devices, Creative Senz3D) operate at 30 samples per second, which is very low compared to the sampling rate of a commodity mouse (default of 100 samples per second). A way to avoid hand tracking is by measuring general movement flow [31], however this technique can easily get confused by other moving objects and body parts in view, which could inject noise.

Due to these limitations we focused our efforts on a system that used markers to track the hand. Note that the use of markers is not inherent to Keyboard Surface Interaction, but only to our particular implementation.

All of our development and evaluation was conducted on a Lenovo laptop computer with a 2.2GHz Intel i7-2670QM processor, 16GB of ram and a screen with 100px per square inch running Fedora 20. The screen resolution was 1366x768 pixels. For all evaluations, we disabled the pointer acceleration feature.

*Sensing*
We used infrared LEDs as markers on the fingers of the user and tracked these LEDs using Wii Remotes (Figure 1), which include tracking capabilities. This approach decreased the overall complexity of the system. For example, position data transmitted by the Wii Remotes required minimal post-processing before this data could be mapped to motion of the pointer on the screen. In contrast, a standard computer vision approach would require image correction, feature extraction, hand recognition and hand tracking. The last two stages are not guaranteed to succeed due to lighting changes and the overall difficulty of hand

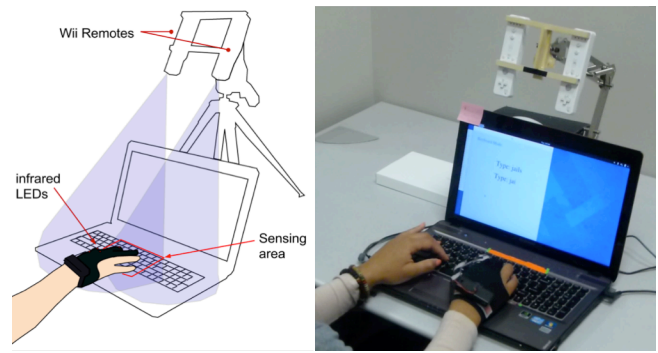

**Figure 1. Basic sensing system. Left: Envisioned system. Right: Actual system and participant during a test**

tracking.

Our simpler and more reliable system allowed us to focus on other aspects of the system like evaluating and optimizing its performance. For tracking the infrared LEDs, we chose Wii Remotes for their price, readily available libraries, infrared tracking capabilities and high sampling rate (100Hz) compared to other infrared systems like the Kinect 1 and 2 (30Hz). Our system had a resolution of 570x300 pixels over a sensing area of 3.3x2.5 inches (~140PPI) shown in Figure 2. This input surface was re-scaled in our software so that every corner on the sensing area would correspond to the corners of the screen. Due to the usage of a high power and high viewing angle infrared LED and the on-board tracking of the wiiRemote, tracking accuracy is 100%.

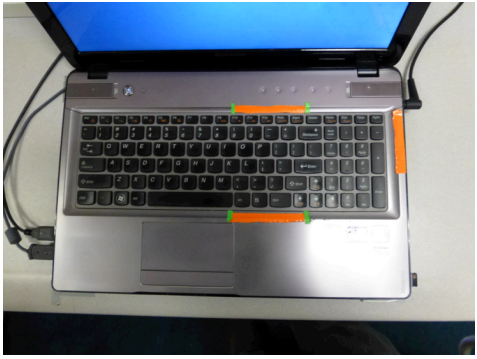

**Figure 2. Fingers sensing area**

We tested two movement-mapping strategies with different versions of the device: *Relative* and *Absolute*. The *Absolute mapping* tracks a single LED located at the tip of the index finger. The movement of this LED is appropriately scaled and translated to move the pointer on the screen directly. This strategy requires the use of only a single Wii Remote.

The *Relative mapping* turns an area of the keyboard into a touchpad; this means that whenever the user touches the surface of the keyboard in that area, she gains control over the pointer. This mapping requires depth estimation to know when the keyboard surface was touched, which was accomplished by using a second Wii Remote and a simplified stereo vision model [23]. Depth estimation also requires some calibration for each user. During this calibration, a participant moves her index finger while touching the desired sensing area over the keyboard. At the same time, our system captures a 3D cloud of points from which the 3D position of the keyboard surface is estimated using least squares regression. The sensing area was roughly 5 by 3 inches, this size was selected because it produced the best tracking, has an aspect ratio close to that of the screen and it was comparable in size to the trackpad's sensing area.

In addition to experimenting with these different mappings, we also built different devices. Initially, and to have minimal equipment wore by our users, we tried using infrared reflecting tape in conjunction with an infrared lamp however the tracking system did not work well with it. The reflecting tape was not diffusing the IR light enough to allow for multiple viewing angles, so at certain angles there was no IR light reflected to the camera and turned into tracking mistakes. In the next few sections, we will describe the different prototypes we tested and the lessons learned from each which influenced the final proof of concept device we used for our evaluations.

### Ring version

The first version of Fingers was composed of a single infrared LED and a small battery (Figure 3 left). This device used absolute mapping. Users performed clicking by moving the ring up and down. While testing the ring ourselves, we found that there was a need to activate and deactivate Fingers. For instance, if the user needs to type and Fingers is active, accidental pointer clicks are likely to occur as the typing motion is similar to the clicking motion. Also, the ring was found to be uncomfortable and the battery life lasted only about 4 minutes making it infeasible for testing with real subjects.

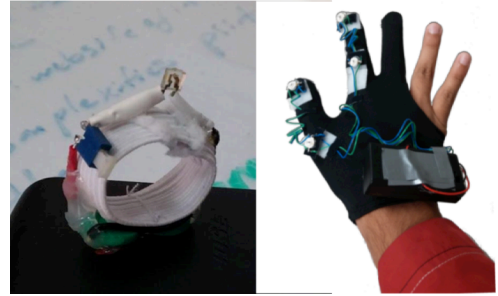

**Figure 3. Left Fingers Ring version, Right Fingers 4 LEDs version**

### 4 LEDs version

To account for the need to activate and deactivate the device, our next version of Fingers introduced a switching gesture. The gesture selected was a pinch with the restriction that it had to be done as shown in Figure 4 right. Making the pinch gesture in a different way, for instance closer to the keyboard or with the hand in a different orientation would not work. When performed, the pinch gesture switched Fingers between typing and pointing mode. For clicking, the gesture in Figure 4 left was used.

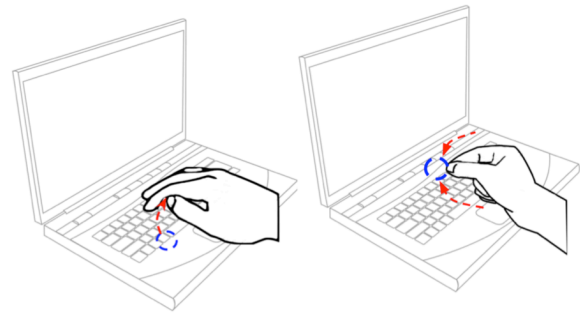

**Figure 4. Left: mouse click gesture; right: mode switch gesture**

To detect these gestures, 4 LEDs were used, with 2 on each (thumb and index) finger (see Figure 3 right). These gestures were first piloted by three of our lab members who found them simple to learn and easy to execute. The idea behind using these gestures was to keep all the controls necessary to execute a pointing task on the tracked hand and to keep the user's hand as close to the keyboard as possible.

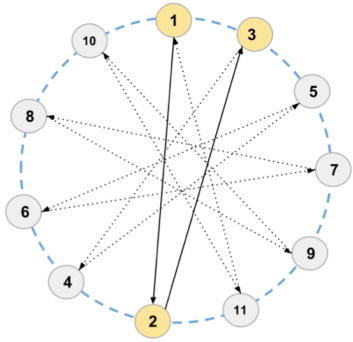

**Figure 5. Standard Fitts' Law test. The user clicks on the targets laid over a circle in the order shown.**

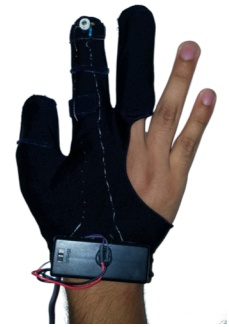

**Figure 6. Final version of Fingers. This version has only one LED located at the tip of the index finger.**

With this version of Fingers, we tried both absolute and relative mapping. In a pilot study, our participants both preferred the relative version and had better performance with it.

Then, using the relative version only, we conducted a study with 30 participants in which they performed a modification of the Fitts' Law Test. This test follows the same structure as in [27] in which targets appears on the screen as shown in Figure 5; however, in addition, in-between consecutive clicks, the participant is asked to type a short word that appears on the screen. A similar test was used in [6] and we refer to this test as the Modified Fitts' Law Test.

*Method*: Each test consisted of 8 blocks. In each block, a different sequence of index of difficulty was presented. For each index of difficulty, there are 11 targets (*i.e.*, 11 clicks) with a fixed inter-target distance of 400 pixels and 10 words of length 4 to 6 characters are shown.

Three different indices (3, 4 and 5, corresponding to three different target sizes) of difficulty (using Shannon's formulation of the index of difficulty [27]) were used, keeping the distance between targets constant as suggested by [12]. To handle order effects we randomized the index of difficulty. Each participant performed the test 3 times, once each for the touchpad, mouse and Fingers. We randomized the order in which participants used these devices. Participants filled out brief questionnaires at the end of the study, aimed at gathering preferences for the different devices.

*Results*: At the end of the study we asked our participants to fill out a small survey to evaluate their experience in the test with the different devices (Fingers, Mouse and Touchpad). We asked open-ended questions like: "What did you like about the <device> and why?", " What did not you like about the <device> and why?". We found that the majority of our participants liked the concept behind Fingers but disliked the gestures describing them as "unnatural", unreliable and tiring. Movement time and homing times were significantly higher than those for the touchpad and mouse. Also, although we found that participants liked the mouse, many complained about having to move their hand back and forth between the mouse and keyboard. Participants had similar complaints with the touchpad, and, overall, the touchpad received more mixed reviews than the mouse.

We concluded that participants did not like to move their hands away from the keyboard. This may be another reason why they did not like to perform the switching gesture, as it required rotating their hand upwards to avoid confusion with typing. For example, without this rotation, pinch-like gestures could be performed accidently when the index finger presses keys near the thumb finger while typing. We also found that the LEDs used in this version of Fingers had too small of a viewing angle (only 60 degrees), which contributed to the Wii Remotes losing track of them often for certain hand poses.

**Final version**
Using the findings from our study we made a number of implementation changes. We replaced the original infrared LEDs with high power (315mW) high viewing angle (140 degrees) infrared LEDs. We also replaced the gestures with key presses for clicking and switching operations. Without any need for gesture recognition we removed all but the LED on the tip of the index finger as can be seen in Figure 6. By using key presses the user's hand will not have to leave the keyboard. This change however brings up the question of which keys would be the best for switching and clicking. For clicking we chose the space bar due to its large size and easy reach. We restricted clicking with the spacebar to be performed only with the hand not being tracked, to avoid mis-clicks caused by clicking and pointing with the same hand. In the next section, we describe how we chose the switching key.

We conducted a series of studies to further optimize the performance of the final version of Fingers. These studies helped us to determine the best performing movement-mapping strategy (relative or absolute) and the fastest keys for switching between typing and pointing mode.

**Absolute *vs*. Relative mapping**
*Method*: To compare Absolute mapping with Relative mapping, we used a standard Fitts' Law test (pointing-only) as described in [27] with a fixed distance of 400 pixels and index of difficulty of 3, 4 and 5 (using Shannon's formulation of the index of difficulty [27]). We had 12 participants (5 males and 7 females) and the whole test took approximately 1 hour to complete. Each participant used Fingers in both the Absolute and Relative mapping modes (we will refer to them as Fingers Absolute and Fingers Relative). For each mapping, the participants went through 8 different blocks of the Fitts' Law test, with each block having a different sequence of indices of difficulty. To handle order effects we randomized the index of difficulty sequences. To handle order effects due to the mapping used (Fingers Relative *vs*. Absolute), half the participants started with Relative mapping, while the other half started with Absolute mapping.

*Results*: After filtering out outliers that were more than 3 standard deviations from the mean, as described in [27], we found that the movement time over the 8 blocks followed the power law of practice for both Fingers Relative ($R^2$=0.911) and Fingers Absolute ($R^2$=0.926), revealing strong learning effects. Due to this effect, only the last block for each device was used to compare movement time between Fingers Relative and Fingers Absolute. The movement time for both Fingers Relative and Absolute does not follow a normal distribution. This was confirmed with a Shapiro-Wilk normality test for Absolute (W=0.83, p-value<0.001) and Relative (W=0.76 p-value<0.001). The median of the movement time across targets clicked was calculated and then the mean across the different indices of difficulty. Fingers Absolute was significantly faster than Fingers Relative (Wilcoxon Signed Rank test p=0.00048, Z=-3.05 and size effect of 0.88). The median movement time was 1.36 seconds for Fingers Relative and 1.13 seconds for Fingers Absolute. Note that this result contradicts the finding of our earlier pilot study; however, the system used for this test is more precise (due to the change in the LEDs used) and is easier to use since it does not rely on gestures. One of the reasons Fingers Relative was slower than Absolute, was that the depth estimation did not work as well when the participant moved her hands quickly. This made the touch detection of the surface of the keyboard(distance of the finger to the surface) noisy, which made it difficult to account for all of the hand movements.

When describing the study, we asked participants to rest their hands on the keyboard; this resulted in their hands being more relaxed and hence trembling and other fatigue artifacts observed in earlier studies were minimized. Most participants initially had difficulty getting used to resting their hand on the keyboard. Most got used to it after the first block and all participants mastered it by the end of the study. Initially, some participants kept their index finger touching the keyboard while keeping their remaining fingers in the "air", much like one uses a touchpad. However, the research supervisor quickly prompted the participants to relax their hand and rest all of the fingers in their natural posture over the keyboard by demonstrating that this did not affect pointer use, unlike a touchpad.

**Switching key**
*Method:* To identify the best key for switching between typing and pointing, we used the Modified Fitts' Law test described earlier. We chose to evaluate the 'alt' key and the 'tab' key. The rationale behind picking these keys is that both are easily reachable (by the left hand's thumb and ring finger, respectively) .For each mapping, the participants went through 8 different blocks of the Fitts' Law test, with each block having a different sequence of indices of difficulty. To handle order effects we randomized the index of difficulty sequences. To handle order effects due to the key used (alt *vs*. tab), half the participants started with the alt key and the other half started with the tab key. For this pilot, we used the same participants as for the previous pilot study. The pilot took approximately 1 hour to complete.

*Results:* We first filtered out outliers using the strategy described in [27]. Then, we found that the homing time for the two keys did not follow a normal distribution, using a Shapiro-Wilk normality test (alt key W=0.73, p-value<0.001 and tab key W=0.75 p-value<0.001). We found that the tab key was significantly faster (Wilcox p-value=0.04 Z=2.01 with a size effect of 0.63). The median homing time was 0.328 seconds for the alt key and 0.269 seconds for the tab key. Note that with this finding, we are not claiming that the tab key is the best key to use for switching but instead that a key in that area of the keyboard would be a good candidate. Our testing software did not use the tab key for typing.

**COMPARATIVE EVALUATION**
Using the final, optimized Fingers, we conducted a comparative evaluation of KSI with Fingers Absolute, the laptop's touchpad and a standard optical Microsoft Mouse (with the acceleration feature deactivated for all devices). The test used was the Modified Fitts' Law test described earlier. For this study we recruited a total of 25 participants: 10 "expert" participants (taken from our earlier pilot studies on Fingers optimization) and 15 novice participants. Our expert group consisted of 4 males and 6 females, with ages ranging from 19 to 42, including 6 students. Expert users had an accumulated experience with Fingers of 32 blocks, 16 on the standard Fitts' Law test and 16 on the Modified Fitts' Law Test. The optimization study and this final evaluation were conducted over two different visits to our lab with the time in between visits being up to 2 weeks. Our novice group consisted of 6 males and 9 females, with ages ranging from 18 to 47, including 11 students.

**Method**
Before beginning the Modified Fitts' Law test, participants filled out a demographic and a discomfort questionnaire (described below). They then had short practice sessions with each of the devices, so they were comfortable in using them. They completed 8 blocks of tests, for each of the three devices. Order effects were minimized by randomizing the order of the index of difficulty sequences in each block, and through using a counterbalanced Latin square, as used in previous pilots. Due to the number of participants, the order was not perfectly balanced. After completing the 8 blocks for a particular device, participants again filled out the discomfort survey, and were given 5 minutes to rest (or longer if needed) to remove any effects of fatigue before starting with the next device. Following the testing with all three devices, participants completed a questionnaire asking about their preferences for each device. Participants received 20 dollars as compensation for their time. The test took between 1 to 1.5 hours to complete.

**Measures**

To measure performance, we looked at homing time and movement time (in seconds), as measured by our instrumented testing software. We also calculated error rate, as the mean of the median number of errors for each index of difficulty for each device and user type. These measures are key components of the performance of each device.

In addition to measuring performance, the study also aimed at measuring fatigue. However, muscle fatigue is, in general, difficult to measure objectively. It requires special equipment to excite a muscle and measure its recovery time after exertion [17]. It also requires sensing equipment capable of measuring the forces exerted by the muscle [17]. Given these complications, we decided to use only subjective measures of fatigue, such as the perceived discomfort rating scale from [18]. In this survey, the participant is asked about her current level of discomfort from 0 (nothing at all) to 10 (Extremely strong) for 6 different parts of the arm commonly used while pointing: hand, wrist, forearm, elbow, upper-arm and shoulder.

Discomfort is calculated by averaging the 6 discomfort rates for the different parts of the arm and subtracting the baseline discomfort rate collected before the study began. The last step was performed to account for any discomfort unrelated to the experiment.

After the study, we asked participants 'what did you like about Fingers and why'? We extracted commonly mentioned words and phrases and grouped those features into themes. For example, we grouped terms such as 'responsive', 'effective', 'smooth', 'fast', and 'precise' into the category of *performance*. Nine participants mentioned performance as a positive of Fingers, nine mentioned movement time, and 6 mentioned reduced fatigue.

**Data Preparation**

In order to pre-process our data, we filtered out outliers as described in [27]. Afterwards, the median homing and movement time was calculated across the different targets was calculated, and then the mean across different indexes of difficulty. We also calculated discomfort across the different devices for expert and novice users.

We tested for learning effect on homing time and movement time. As expected, novice users' performance showed a learning effect while experts did not. Homing time for novice users fit the power law of learning for Fingers ($R^2=0.94$), had no effect for the trackpad ($R^2=0.31$), and some effect for the mouse ($R^2=0.71$). None of the other metrics showed a learning effect. Thus, for novice users we only looked at the last block for homing time of fingers and the mouse, and computed the average across blocks for the trackpad.

A Shapiro-Wilk normality test showed that none of our movement time data followed a normal distribution, at a level α=0.05. Because the data was not normal, we used a Wilcoxon signed rank test in each case to measure whether there was a significant difference between the devices for each metric. All significance values presented are from these comparisons.

Next we present an analysis of the results on a metric-by-metric basis. Because there was no significant difference in the error rates across the devices (which varied from 0.033 to 0.09), we will not discuss error rate further. The median scores for homing time, movement time, and discomfort are shown in Figure 7. Bars with an *e* show expert performance while bars with an *n* show novice performance.

**Homing time**

Fingers was about a tenth of a second faster than the mouse and trackpad, for experts, and novices achieved expert homing performance by the final block of the study. Expert median homing time was 0.23s, while mouse and trackpad both required over 0.31s (a significant difference: Z=-2.7 for trackpad; Z=-1.9 for mouse; p<.01 in both cases). For novice users, the learning effect for homing time was strong (power law of practice with $R^2=0.94$) and their performance was 0.27s for Fingers, 0.35s for the trackpad and 0.37s for the mouse by the final block (the block we will present the results from), as can be seen in Figure 7. As mentioned earlier, experts did not have a learning effect on any devices for homing time, thus we conclude that Fingers provides the best homing time.

When we examine in detail the distribution of homing time for the touchpad and the mouse, we found that there was an unusual peak centered over the zero bin in our histogram. This explains in part why the homing time for the mouse and touchpad does not follow a normal distribution. In the discussion section, we will describe why the peak occurred.

As described in the measures, homing time in Fingers is essentially the time required to switch modes. When we asked participants what they disliked about Fingers, most participants (14/25) did not like to use the tab key to switch between typing and pointing. Thus, there may be further opportunity to improve homing time with an alternative key.

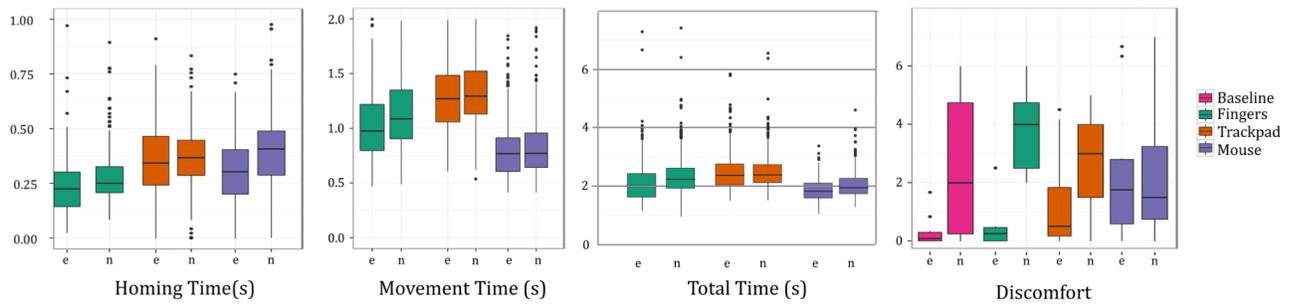

Figure 7: Box plots for the different performance metrics across devices, Novice (n) and Expert (e) users. The different units are: Time in seconds and discomfort total average score from 0 to 10.

**Movement time**

In the case of movement time, not surprisingly, the mouse was the best among the three devices, with a median time of 0.78s and 0.73s (for novices and experts, respectively). In contrast, Fingers had median movement time of 1.18 for novices and 0.96s for experts and the Trackpad had a time of 1.41 for novices and 1.33s for experts. These results are summarized in Figure 7, and the difference between the mouse and Fingers is significant (Z=3.4 for novices and Z=2.8 for experts, p<.001 in both cases). Unlike with homing time for novice users, there was no learning effect.

**Table 1. Total movement time across devices**

|  | Fingers |  | Trackpad |  | Mouse |  |
| --- | --- | --- | --- | --- | --- | --- |
|  | Novice | Expert | Novice | Expert | Novice | Expert |
| Total time | 1.72s | 1.42s | 2.11s | 2.01s | 1.52s | 1.35s |

We also calculated the *median total movement time*, as the sum of homing time + movement time + return time. We define return time as the time elapsed when typing starts after the user has clicked on the on-screen target. As shown in Figure 7, the total movement time for experts is comparable to that of the mouse (within 9%). As shown in Table 1, Fingers had a median total movement time of 1.72s for novices and 1.42s for experts, the trackpad 2.11s for novices and 2.01s for experts and the mouse 1.52s for novices and 1.35s for experts.

**Discomfort rate**

As is visible in Figure 7, novices did not report any significant difference in discomfort across devices. Interestingly, in the post-study survey, they perceived both Fingers and the mouse as causing less fatigue (Figure 8, left).

In contrast, expert participants rated both Fingers (median discomfort=0.0) and the trackpad (median discomfort=0.4) as causing significantly less discomfort than the mouse (average discomfort=1.5) (Z=-2.3, p<.01 in both cases). In the post-study survey, expert users favored Fingers, rated the trackpad as neutral and rated the mouse as causing the most fatigue. Figure 8 shows the number of participants rating each device as less, neutral, or more fatiguing.

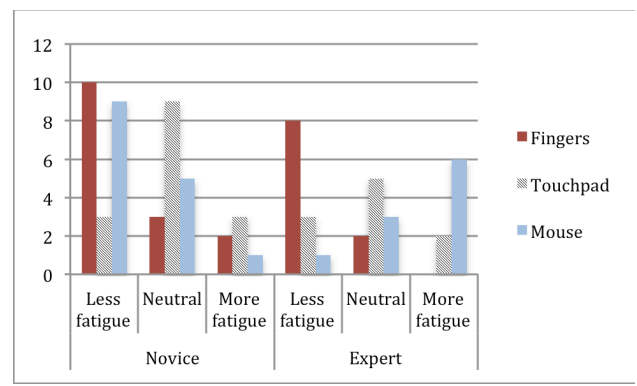

Figure 8. Summary of responses to the question: Please rate the level of fatigue induced by the device in this test.

**DISCUSSION AND FUTURE WORK**

Since the learning curve was fairly rapid, and only affected homing time, we focus our discussion of the results on expert use. In expert use, Fingers outperformed the trackpad on every measure and outperformed the mouse on homing time and discomfort. Although it performed worse than the mouse on movement time, its gains in homing time allowed it to come within 9% of mouse performance on total movement time, with no increase in error rate. This performance comes at a relatively cheap cost (less than US$50 for our relatively simple final prototype).

As mentioned in the results section, we observed some homing times of zero for the touchpad and mouse. This observation can be explained by some of our participants' behavior: In these cases, the participant reaches for the trackpad or mouse (and moves the pointer) even while still typing with the left hand. This only occurred if the last letters of the word on the screen to be typed were located on the left part of the keyboard. This behavior had a higher frequency for experts (3% of all blocks) than for novice users (1% of all blocks).

One unexpected result is that novice users did not perceive a difference in discomfort between Fingers and the mouse. We hypothesize that expert users learned to use Fingers in a more optimal way and this is why for novice users there was no significant difference in discomfort rates.

As we learned from our qualitative survey, our participants did not like using the tab key for switching. This was surprising since the quantitative results for homing time are very good in general. The main problem with the switching key was that it was confusing to use given its regular use for indenting while typing on daily computing tasks. This problem could be solved in three different ways: 1) Creating a custom key for switching on KSI devices to avoid confusion. 2) Using the tab key to switch only from typing to pointing. To switch from pointing to typing, the user can simply start typing. We did not explore this possibility as we thought it would be confusing and distracting to see the pointer moving on its own during the switch from typing to pointing. This could be further explored and evaluated. 3) Eliminating completely the need for a switch. This may be possible by detecting dragging of the hand over the keyboard surface. This could not be achieved with Fingers due to our limited sampling rate, which reduces the precision of our depth estimation.

While participants did not complain in our surveys about the glove, the use of the glove on which the LED and battery are mounted, could be a problem over the long run due to wear and tear, weight caused by the battery pack and heat from wearing a glove. Also it would force the users to wear something every time they want to perform a pointing task even for short-lived tasks. One way to solve this problem could be to use a high-speed camera and very robust hand tracking technology.

**Implementing future KSI devices**

One fundamental drawback of our proof of concept for becoming a commercial product was the usage of a glove. This could be avoided in the future using robust hand tracking and improved depth sensing technologies. For example, [28] shows robust hand tracking using a Kinect2 at 120fps (higher than our device) and at 512x424 pixels (lower resolution than ours). This fast and accurate hand tracking technology can be used directly to produce a KSI device. We expect similar improvements in depth sensing to make our approach field-ready. In addition, multi-device set ups (e.g., watch plus smart phone) are becoming increasingly common, and it is not unrealistic to expect some sort of hand/wrist worn device (or external sensor, similar to the Leap) that could help make KSI a reality. Another aspect of improvement over our own implementation would be a switch-free KSI device: a modality in which the KSI device knows the intention for a pointing task without any explicit gesture or activation. An example of this approach is demonstrated in [30] where a touch sensitive layer on top of the keyboard can recognize pointing intention, although [30] is too slow for a KSI device, further improvements could make homing time almost zero and this device would be even faster than the mouse. For example, assuming a homing time for Fingers for both novice and expert users of 0.1 seconds and keeping movement time the same, Fingers becomes 9% faster for novice users and 14% faster for expert users when compared with the mouse.

Another important factor to consider is the specifications of the sensing technology used to implement a KSI device. With Fingers, we had a spatial resolution of 140PPI, 100 samples per second and 0 error rate (100% tracking accuracy). Any further development of a KSI device should use these specifications as a baseline for obtaining similar results.

In our evaluation of Fingers relative *vs*. absolute, we found that the absolute mapping was the fastest; however, this mapping has a shortcoming: It cannot make use of acceleration features. This could potentially make the absolute mapping slower when compared to the touchpad and the mouse with acceleration activated. There are two additional issues to consider in our future work 1) integrating the acceleration feature into Fingers' relative mapping. 2) While we conducted our final evaluation with Fingers absolute, since our participants were resting their hands on the keyboard throughout the experiment (like they would do using Fingers Relative), we would expect an improved version of Fingers relative to also work well.

Last, with Fingers we have shown that KSI is promising, and we imagine a future in which we no longer need a touchpad or a mouse. This, would free up space and improve the human-computer experience. With expected advances in depth sensing and hand tracking, KSI devices with similar performance that do not need of wearing a globe or any trackers can be expected to rapidly become available.

KSI also has its limitations. It relies heavily on the smoothness of the surface of the keyboard. For example, KSI is not a good option for keyboards that have a rougher surface, or keys that have deep depressions in them, as users' fingers could get "stuck", slowing down performance. Nonetheless, on current laptop and some desktop keyboards, KSI works well.

**CONCLUSIONS**

In this paper, we introduced the concept of Keyboard Surface Interaction, and evaluated it through Fingers, our proof of concept. Fingers has better performance (with expert users) than that of the touchpad for both time to perform pointing actions and perceived comfort levels. When compared to the mouse, it has mixed performance with respect to the time to perform pointing actions, but has better perceived comfort levels. In addition, Fingers had the same error rate as both the trackpad and the mouse.

This means that KSI devices can be a very good alternative to the trackpad, and potentially even the mouse, if comfort is a priority. Furthermore, we have shown that the surface of the keyboard can be used to do much more than has been explored up to date. In this paper, we only explored the execution of pointing tasks, but the potential is there to use the keyboard surface as a multi-touch sensing device on which to scroll, zoom and perform touch gestures, similar to a touchpad or a touchscreen.